# Nuclear physics inputs needed for geo-neutrino studies

**G Bellini[1,2], G Fiorentini[3,4], A Ianni[5], M Lissia[6], F Mantovani[4,7,9] and O Smirnov[8]**

[1] Dipartimento di Fisica, Università degli Studi di Milano, I-20100 Milano, Italy

[2] Istituto Nazionale di Fisica Nucleare, Sezione di Milano, I-20100 Milano, Italy

[3] Dipartimento di Fisica, Università degli Studi di Ferrara, I-44100 Ferrara, Italy

[4] Istituto Nazionale di Fisica Nucleare, Sezione di Ferrara, I-44100 Ferrara, Italy

[5] INFN, Laboratori Nazionali del Gran Sasso, I-67010 Assergi (AQ), Italy

[6] Istituto Nazionale di Fisica Nucleare, Sezione di Cagliari, I-09042 Monserrato (CA), Italy

[7] Centro di GeoTecnologie CGT, I-52027 San Giovanni Valdarno (AR), Italy

[8] Joint Institute for Nuclear Research, 141980, Dubna, Moscow region, Russia

**Abstract**. Geo-neutrino studies are based on theoretical estimates of geo-neutrino spectra. We propose a method for a direct measurement of the energy distribution of antineutrinos from decays of long-lived radioactive isotopes.

## 1. Introduction

Geo-neutrinos, the antineutrinos from the progenies of U, Th and $^{40}K$ decays in the Earth, bring to the surface information from the whole planet, concerning its content of radioactive elements. Their detection can shed light on the sources of the terrestrial heat flow, on the present composition, and on the origins of the Earth.

Although geo-neutrinos were conceived very long ago, only recently they have been considered seriously as a new probe of our planet, as a consequence of two fundamental advances that occurred in the last few years: the development of extremely low background neutrino detectors and the progress on understanding neutrino propagation. From the theoretical point of view, the links between the geo-neutrino signal and the total amount of natural radioactivity in the Earth have been analyzed by several groups; various "reference models" for geo-neutrino production have been presented in the literature; these have been refined with geological and geochemical studies of the region surrounding the detector. KamLAND and Borexino are collecting geo-neutrino data, while several planned experiments (e.g. SNO+, LENA, HANOHANO, EARTH…) have geo-neutrino measurements among their primary goals. A recent review is presented in [1].

[9] Corresponding author: fabio.mantovani@unisi.it

It seems to us that this most intense and most interesting activity has to be complemented with some deepening of the nuclear physics which is at the basis of geo-neutrino production and which is crucial for interpreting future geo-neutrino data.

All experiments, either running or in preparation, use the well tagged inverse beta decay on free protons

$$\bar{\nu}_e + p \rightarrow e^+ + n - 1.806 \text{ MeV} \tag{1.1}$$

as the reaction for geo-neutrino detection. The signal is estimated from the cross section $\sigma(E_\nu)$ of (1.1) and from the decay spectrum $f(E_\nu)$ of geo-neutrinos produced in the decay chains of $^{235}U$ and $^{232}Th$, the relevant quantity being the "specific signal" defined as:

$$s = \int_{E_0}^{E_{MAX}} dE_\nu \sigma(E_\nu) f(E_\nu) \tag{1.2}$$

where the integral is between the threshold energy ( $E_0 = 1.806$ MeV ) and the maximal geo-neutrino energy $E_{MAX}$ and the spectrum $f(E_\nu)$ is normalized – say - to the number N of geo-neutrinos produced in each decay chain, $N = \int_0^{E_{MAX}} dE_\nu \, f(E_\nu)$.

It is important to observe that the specific signal is affected by unknown uncertainties. In fact, whereas the cross section $\sigma(E_\nu)$ is known with an accuracy of half of per cent or better, it is difficult to assess the accuracy of $f(E_\nu)$, which is determined from rather indirect measurements and questionable theoretical assumptions.

*The goal of this note is to provide a framework for a direct measurement of* $f(E_\nu)$*, so that the accuracy of the specific signal can be established.*

## 2. Why should geo-neutrino spectra be measured?

Geo-neutrinos are produced through β and βγ processes:

$$X \rightarrow X' + e + \bar{\nu}_e \tag{1.3}$$

$$X \rightarrow X'^* + e + \bar{\nu}_e, \quad \hookrightarrow X' + n\,\gamma \tag{1.4}$$

In order to determine the geo-neutrino decay spectra $f(E_\nu)$ one has thus to know:

i. the probabilities of populating the different energy states of the final nucleus;
ii. the shape of the neutrino spectrum for each transition.

Probabilities are generally derived by measuring the relative intensities of gamma lines. In particular, the transition to the ground state (pure $\beta$) is indirectly determined by a subtraction procedure, which brings with it the propagation of all errors on the measured intensities. We remark that this transition, which is associated with the most energetic geo-neutrinos, gives the largest contribution to the specific signal.

For each transition, the shape of the neutrino spectrum is generally calculated assuming the well known "universal shape". This expression, see [1], corresponds to momentum independent nuclear matrix elements (as for allowed transitions) and includes the effect of the bare Coulomb field of the nucleus through the relativistic Fermi function. Electron screening and nuclear finite size effects are not considered. Furthermore, the same "universal shape" expression is used for forbidden transitions, where momentum dependent nuclear matrix element can appear.

All this suggests to us that some experimental check is needed.

## 3. Towards a direct measurement of geo-neutrino decay spectra

When the nucleus $X$ decays, whichever is the transition involved, energy conservation provides a connection between the neutrino energy $E_\nu$, the kinetic energy of the electron $T_e$ and the total energy of the emitted gammas, $E_\gamma$ :

$$Q = E_\nu + T_e + E_\gamma \qquad (1.5)$$

where $Q = M_X - M_{X'} - M_e$ is the $Q$-value for the decay.

In order to measure the geo-neutrino spectrum, one needs a detector which is capable of measuring the "prompt" energy deposited together by the electron and gammas, $E_{prompt} = T_e + E_\gamma$. When measured decay events are displayed as a function of $E_{prompt}$, by a specular reflection one immediately obtains the number of events as a function of neutrino energy, at $E_\nu = Q - E_{prompt}$, see Figure 1 for the decay spectrum of $^{214}$Bi.

For such a measurement one needs a detector that can collect the energy lost by both electrons and gammas, and which has a similar response to both electrons and gammas. Essentially, this is a calorimetric measurement. In principle, it can be done with bolometers, which have very good energy resolution but long dead times. A sufficiently large liquid scintillator detector is suitable for such measurements. Although energy resolution is limited, nevertheless it can contain both electrons and gammas and a significant statistics can be collected in a short time.

We suggest to exploit the potential of the Counting Test Facility (CTF), which is operational and available in the underground Gran Sasso Laboratory.

Like Borexino, the CTF design [2] is based on the principle of graded shielding, see Figure 2 and 3. The active scintillation liquid in CTF is a 4 ton mass of pseudocumene enclosed in a transparent nylon sphere, the CTF vessel. Outside this vessel there is a volume of ultra-pure water which is enclosed in a second nylon sphere, the CTF shroud. A set of inward-facing PMTs is arrayed outside the shroud. The entire apparatus, surrounded by another volume of water, is contained in a cylindrical stainless tank. The bottom surface of the tank holds 16 upward-facing PMTs used to tag the tracks of muons passing through the detector.

The facility is equipped with a rod system, which can be used to insert a small, cylindrical quartz vial inside the CTF vessel. A suitable source, dissolved in the liquid scintillator, can be placed in the vial. Electrons are stopped inside the vial and the scintillation light is propagated within CTF through the quartz (which is transparent to the near-UV wavelengths of scintillation light and has index of refraction near that of the scintillator) whereas gamma conversion occurs inside the CTF inner vessel. The inward facing PMTs outside the shroud can thus detect light originating from both electrons and gammas.

We think that CTF, in addition to being available and operational, offers a particular advantage. The detection method which we propose for measuring geo-neutrino decay spectra is rather similar to that which will be adopted by all experiments (KamLAND, Borexino, SNO+…) for detection of geo-neutrino interactions. In particular, Borexino has the same liquid scintillator and the same phototubes as CTF. Some systematic errors will cancel when extracting the geo-neutrino interaction signal in Borexino by using the geo-neutrino decay spectrum measured with CTF.

## 4. What has to be measured?

Geo-neutrinos with energy above the threshold for (1.1) arise only from the chains of $^{238}$U and $^{232}$Th.

In particular, for $^{238}$U only three nuclides ($^{234}$Pa, $^{214}$Bi, $^{210}$Tl) contribute to the geo-neutrino signal. The contribution from $^{210}$Tl is negligible, due to its small occurrence probability, and the uranium contribution to the geo-neutrino signal comes from five $\beta$ decays: one from $^{234}$Pa and four from $^{214}$Bi. Table 1 lists the effective transitions, i.e., those that can produce antineutrinos with energy above the threshold $E_0$. In fact, 98% of the uranium signal arises from the two transitions to the ground state (in bold in Table 1) and an accuracy better than 1% is achieved by adding the third one.

$^{232}$Th decays into $^{208}$Pb through a chain of six α decays and four $\beta$ decays. In secular equilibrium the complete network includes five $\beta$-decaying nuclei. Only two nuclides ($^{228}$Ac and $^{212}$Bi) yield

antineutrinos with energy larger than 1.806 MeV. The thorium contribution to the geo-neutrino signal comes from three $\beta$ decays: one from $^{212}$Bi and two from $^{228}$Ac (see Table 2). In fact, 94% of the thorium signal arises from the transition to the ground state of $^{212}$Po (in bold in Table 2).

We remind that, assuming the chondritic ratio for the global uranium and thorium mass abundances, $a(Th)/a(U)=3.9$, one expects that geo-neutrinos from uranium (thorium) contribute about 80% (20%) of the total U+Th geo-neutrino signal.

In summary:

*a) 98% of uranium geo-neutrino signal comes from just two transitions, one from $^{214}$Bi and the other from $^{234}$Pa. They provide 77% of the expected total U+Th signal.*

*b) A single decay of $^{212}$Bi accounts for 94% of the thorium signal. It provides 20% of the expected U+Th signal.*

Just three transitions have to be investigated experimentally. In this respect, the following considerations can be useful:

a) $^{222}$Rn ($\tau_{1/2}$ = 3.8 days) can be easily dissolved in the scintillator and the decay of $^{214}$Bi is uniquely identified by the subsequent decay of $^{214}$Po ($\tau_{1/2}$ = 0.16 ms)

b) By dissolving $^{238}$U in the scintillator, one can detect the beta decay of $^{234}$Pa (superimposed however with that of $^{234}$Th). The subsequent decays of the chain are effectively blocked by the long half-live of $^{234}$U ($\tau_{1/2}$ = 245.5 Kyr ).

c) For the investigation of $^{212}$Bi decay one has to start with a $^{224}$Ra source ($\tau_{1/2}$ = 3.7 days) or with a $^{232}$Th source. The decay of $^{212}$Bi can be easily identified by the subsequent $\alpha$-decay of $^{212}$Po ($\tau_{1/2}$ = 0.3 μs).

An experimental program on these lines has already started at Gran Sasso.

**Acknowledgments**

We are grateful for enlightening discussions and valuable comments to L. Miramonti, B. Ricci and C. Tomei.

This work was partially supported by MIUR (Ministero dell'Istruzione, dell'Università e della Ricerca) under MIUR-PRIN-2006 project "Astroparticle physics".

## References

[1] Fiorentini G, Lissia M and Mantovani F 2007. Geo-neutrinos and earth's interior, *Phys. Rep.*, **453/5-6** 117-172.

[2] Borexino Collaboration, 1998. A large-scale low-background liquid scintillation detector: the counting test facility at Gran Sasso, *NIM*, **406** 411.

**Table 1.** Effective transitions in the $^{238}$U chain, from [1]

| i→j | $R_{i,j}$ | $E_{MAX}$ [keV] | $I_k$ | $\Delta I_k$ | Type | $S_U$ [%] | $S_{tot}$ [%] |
|---|---|---|---|---|---|---|---|
| **$^{234}Pa_m \rightarrow ^{234}U$** | **0.9984** | **2268.92** | **0.9836** | **0.002** | **1st forbidden $0^- \rightarrow 0^+$** | **39.62** | **31.21** |
| **$^{214}Bi \rightarrow ^{214}Po$** | **0.9998** | **3272.00** | **0.182** | **0.006** | **1st forbidden $1^- \rightarrow 0^+$** | **58.21** | **45.84** |
| | | 2662.68 | 0.017 | 0.006 | 1st forbidden $1^- \rightarrow 2^+$ | 1.98 | 1.55 |
| | | 1894.32 | 0.0743 | 0.0011 | 1st forbidden $1^- \rightarrow 2^+$ | 0.18 | 0.14 |
| | | 1856.51 | 0.0081 | 0.0007 | 1st forbidden $1^- \rightarrow 0^+$ | 0.01 | 0.01 |

**Table 2.** Effective transitions in the $^{232}$Th chain, from [1].

| i→j | $R_{i,j}$ | $E_{MAX}$ [keV] | $I_k$ | $\Delta I_k$ | Type | $S_{Th}$ [%] | $S_{tot}$ [%] |
|---|---|---|---|---|---|---|---|
| **$^{212}Bi \rightarrow ^{212}Po$** | **0.6406** | **2254** | **0.8658** | **0.0016** | **1st forbidden $1^- \rightarrow 0^+$** | **94.15** | **20.00** |
| $^{228}Ac \rightarrow ^{228}Th$ | 1.0000 | 2069.24 | 0.08 | 0.06 | Allowed $3^+ \rightarrow 2^+$ | 5.66 | 1.21 |
| | | 1940.18 | 0.008 | 0.006 | Allowed $3^+ \rightarrow 4^+$ | 0.19 | 0.04 |

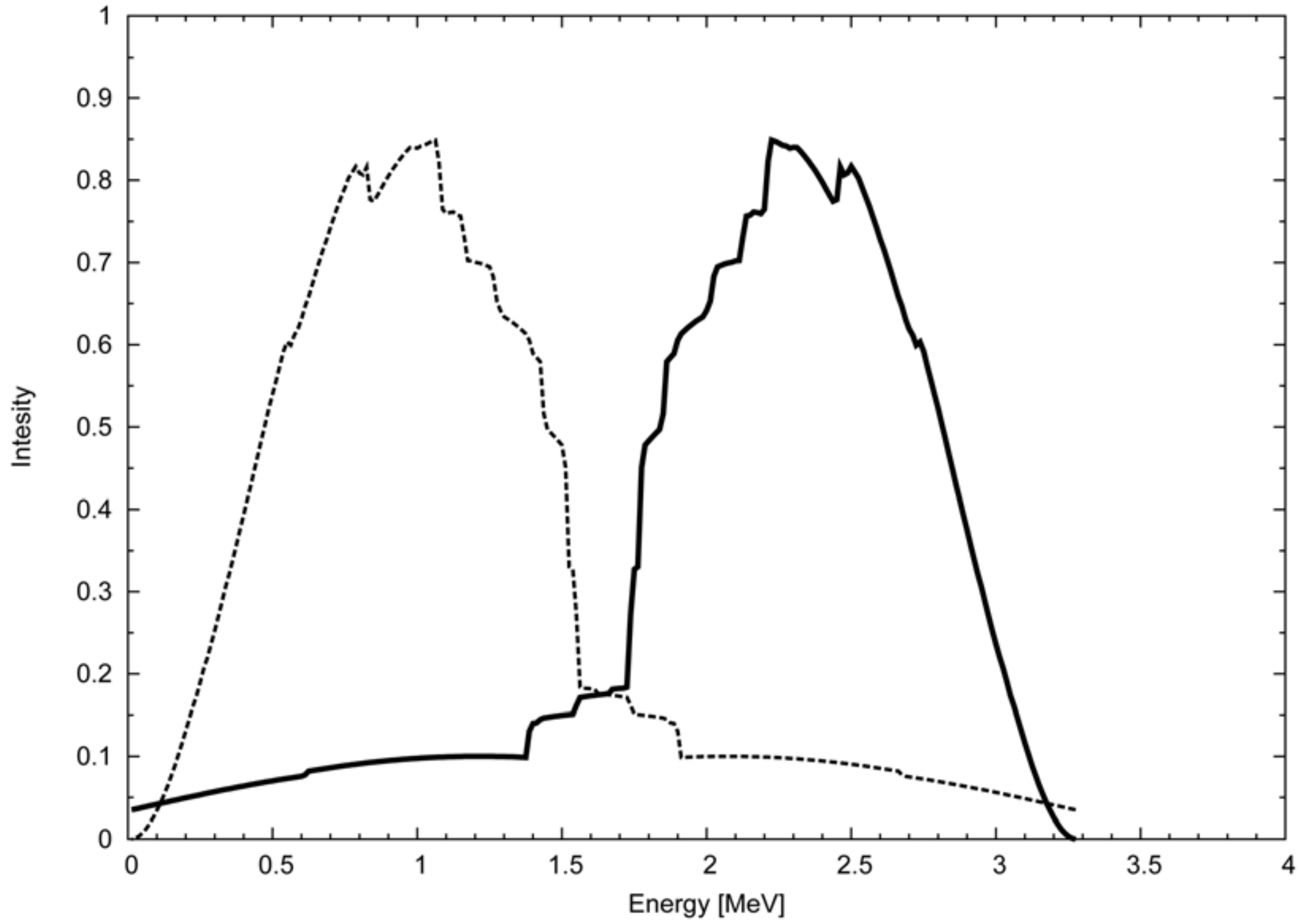

**Figure 1.** The decay spectrum of $^{214}$Bi as a function of the prompt energy $E_{\rm prompt}$ (full line) and of the neutrino energy $E_\nu$ (dashed line).

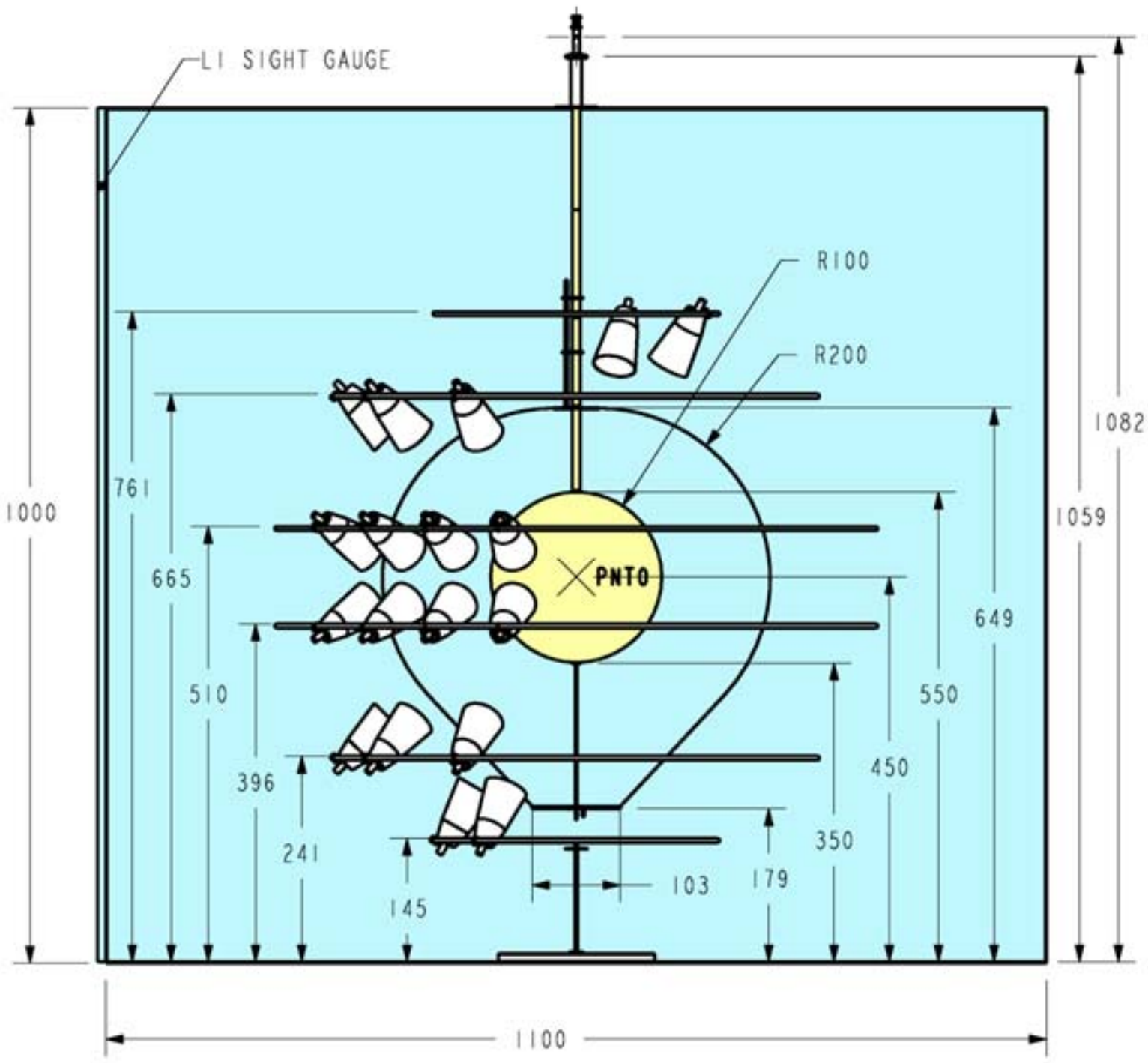

**Figure 2.** Side view of the design of CTF. The vessel (labelled R100 in this drawing) and shroud (R200) are shown, as well as the six rings of PMTs, the cylindrical tank, and the tubes used for filling and draining the vessel. The point PNT0 is the nominal center of the sphere of PMTs and of the CTF vessel. Dimensions are given in cm. Courtesy of the Borexino collaboration.

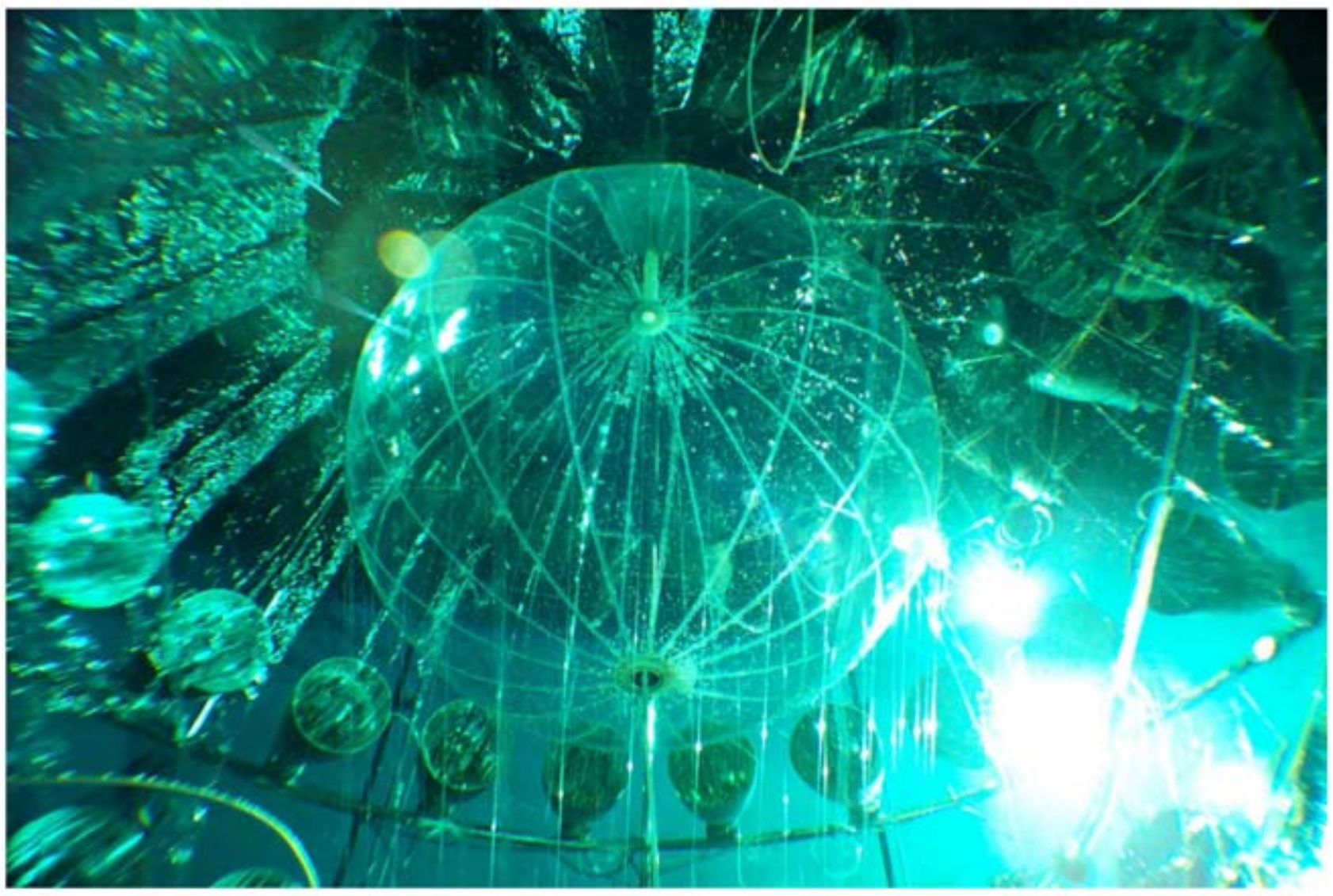

**Figure 3.** A photograph of the CTF viewed from below. Courtesy of the Borexino collaboration..